\documentclass[final,5p,times,twocolumn]{elsarticle}

\usepackage{epsfig}

\usepackage{amssymb}
\journal{}

\begin{document}

\begin{frontmatter}

%% Title, authors and addresses

%% use the tnoteref command within \title for footnotes;
%% use the tnotetext command for theassociated footnote;
%% use the fnref command within \author or \address for footnotes;
%% use the fntext command for theassociated footnote;
%% use the corref command within \author for corresponding author footnotes;
%% use the cortext command for theassociated footnote;
%% use the ead command for the email address,
%% and the form \ead[url] for the home page:
%% \title{Title\tnoteref{label1}}
%% \tnotetext[label1]{}
%% \author{Name\corref{cor1}\fnref{label2}}
%% \ead{email address}
%% \ead[url]{home page}
%% \fntext[label2]{}
%% \cortext[cor1]{}
%% \address{Address\fnref{label3}}
%% \fntext[label3]{}

\title{Quantum Reflection and Transmission 
in Ring Systems with Double Y-Junctions: 
Occurrence of Perfect Reflection} 

%% use optional labels to link authors explicitly to addresses:
%% \author[label1,label2]{}
%% \address[label1]{}
%% \address[label2]{}

\author[oita]{Yukihiro Fujimoto}
\ead{y-fujimoto@oita-ct.ac.jp}

\author[toma]{Kohkichi Konno}
\ead{kohkichi@tomakomai-ct.ac.jp}

\author[toma]{Tomoaki Nagasawa}
\ead{nagasawa@tomakomai-ct.ac.jp}

\author[toma]{Rohta Takahashi}
\ead{takahashi@tomakomai-ct.ac.jp}

\address[oita]{National Institute of Technology, Oita College, 
             1666 Maki, Oita 870-0152, Japan}
\address[toma]{National Institute of Technology, Tomakomai College, 
             443 Nishikioka, Tomakomai 059-1275, Japan}

\begin{abstract}
%% Text of abstract
We consider the scattering problems of a quantum particle 
in a system with a single Y-junction and in ring systems
with double Y-junctions. We provide new formalism for
such quantum mechanical problems.
Based on a path integral approach, we find compact
formulas for probability amplitudes in the ring systems.
We also discuss quantum reflection and transmission 
in the ring systems under scale-invariant junction conditions.
It is remarkable that perfect reflection can occur 
in an anti-symmetric ring system, in contrast with 
the one-dimensional quantum systems having singular nodes
of degree 2. 
\end{abstract}

\begin{keyword}
%% keywords here, in the form: keyword \sep keyword

one-dimensional quantum wires \sep Y-junctions 
\sep transmission \sep reflection  

%% PACS codes here, in the form: \PACS code \sep code

\PACS 03.65.-w \sep 03.65.Xp \sep 03.65.Db

%% MSC codes here, in the form: \MSC code \sep code
%% or \MSC[2008] code \sep code (2000 is the default)

\end{keyword}

\end{frontmatter}

%% \linenumbers

%% main text
\section{Introduction}
\label{sec1}

It is truly interesting that the variety of 
junction conditions appears when the number 
of space dimension is reduced in quantum mechanics. 
This feature becomes apparent in one-dimensional quantum systems.
For example, when we consider a node 
(i.e., a point interaction)
on a one-dimensional quantum wire, 
the two adjacent one-dimensional spaces are completely 
separated by the node. This situation 
leads to various non-trivial junction conditions. 
The characteristics of the non-trivial junction conditions 
in one-dimensional quantum systems have comprehensively been 
studied by mathematical works \cite{rs,seba,aghh,cft}.

We shed light on quantum problems in a system of a Y-junction. 
The Y-junction is composed of 
three one-dimensional quantum wires. These wires
intersect at one point. The intersection of degree 3 has a 
point interaction parametrized by U(3) \cite{rs,seba,aghh,cft}.
Such a system was originally investigated by pioneer works 
(e.g., \cite{bia,buttiker})
with a simplified $S$-matrix, which was inspired by realized nano-rings.
Some mathematical features of transmission of a quantum particle in the system 
with a Y-junction were investigated in \cite{cet} 
and also in the context of a star graph etc \cite{tm,aj,ohya,tc}. 
However, the thorough investigation of 
the system has not yet been completed.

In the present work, we provide new formalism 
for the quantum mechanical problems in the system with a single 
Y-junction and ring systems with double Y-junctions.
In particular, by focusing on scale-invariant 
junction conditions, we investigate quantum reflection 
and transmission in such systems.  
This paper is organized as follows.
In \S \ref{sec2}, we formulate the system with a
single Y-junction. We also formulate the
ring systems with double Y-junctions in \S \ref{sec3}. 
In \S \ref{sec4}, we discuss quantum reflection and 
transmission on the ring systems under the scale-invariant
junction conditions. 
Finally, we provide a summary in \S \ref{sec5}.

\section{Formulation of a system with a single Y-junction}
\label{sec2}

\subsection{The Schr\"odinger equation}

\begin{figure}[tbh]
  \includegraphics[width=.45 \textwidth]{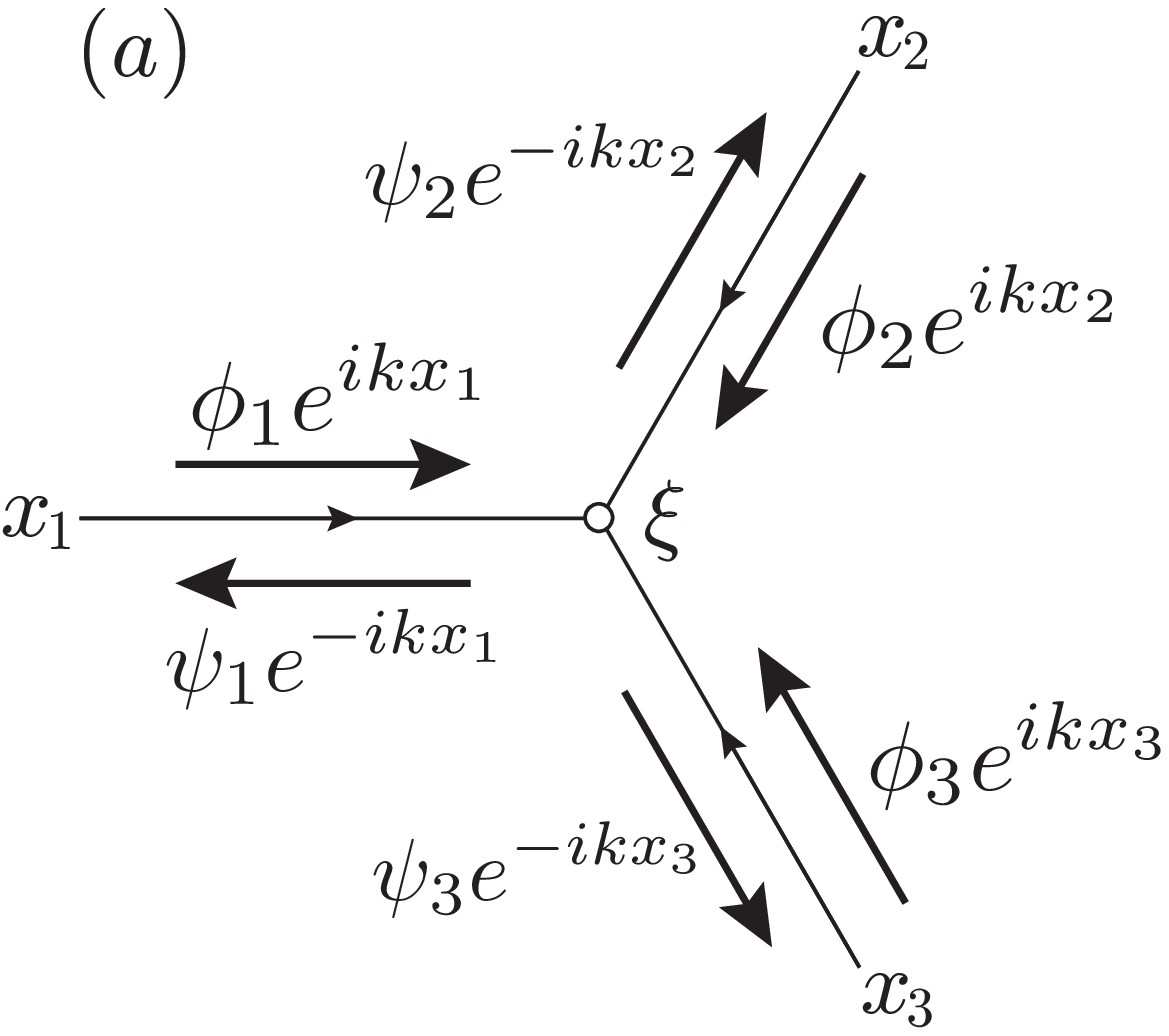}
  \includegraphics[width=.45 \textwidth]{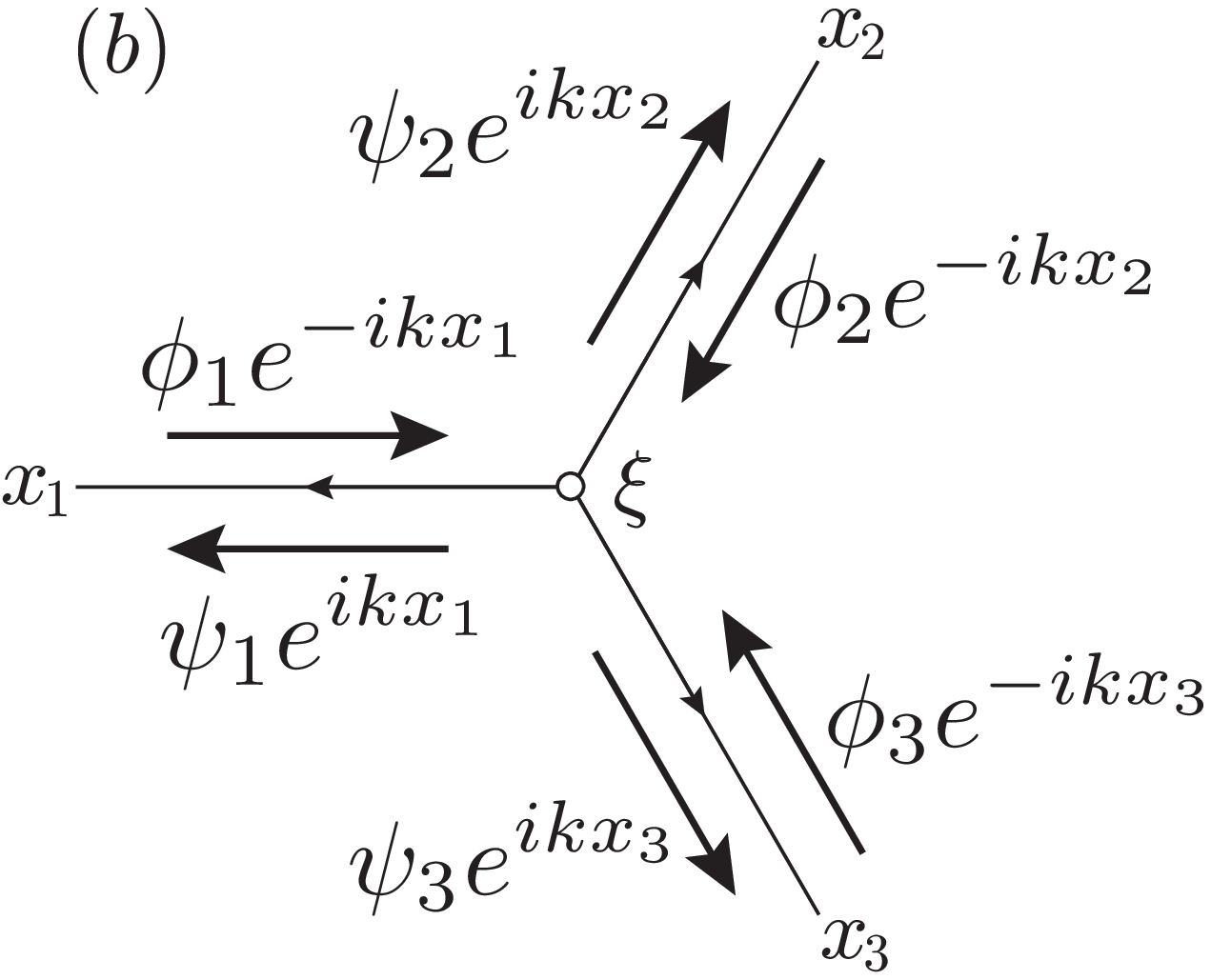}
  \caption{A quantum system with a Y-junction.  
          The tree axes are labeled by $x_{1}$, $x_{2}$ and $x_{3}$.
          The node is given by $x_{i}=\xi \ (i=1,2,3)$. 
          (a) All axes are directed to the node.
          (b) The axes are taken in the opposite direction
           in comparison with (a).} 
  \label{fig1}
\end{figure}

We consider a quantum system with a single Y-junction, 
in which three one-dimensional quantum wires intersect at one point. 
Let us assume that the three axes are given 
by $x_{1}$, $x_{2}$ and $x_{3}$ and 
directed to the node as shown in Fig.~\ref{fig1}(a)
(inward axes).
We also assume that the node locates 
at $x_{i} = \xi \ (i=1,2,3)$.
Note that the angle between any two axes has no effect 
on the physical states. 
On each wire ($x_{i} < \xi$), 
a quantum particle with mass $m$
obeys the Schr\"odinger equation
\begin{equation}
\label{eq:schrodinger}
 i \hbar \frac{\partial}{\partial t} \Phi_{i} \left( t, x_{i} \right)
 =  - \frac{\hbar^2}{2m} \frac{\partial^2}{\partial x_{i}^2}
   \Phi_{i} \left( t, x_{i} \right) ,
\end{equation}
where $\Phi_{i}$ denotes the wave function on the $x_{i}$-axis.
Thus we assume a free particle on the wire.

\subsection{Junction conditions}

Let us discuss the expression of a junction condition.
The junction condition at the node 
is provided by the conservation of the probability current
\begin{equation}
\label{eq:c-conserv}
 j_{1}(t, \xi)  + j_{2}(t, \xi) + j_{3}(t, \xi) =0 ,
\end{equation}
where the probability current $j_{i} (t, x_{i})$ on the $x_{i}$-axis 
is given by 
\begin{equation}
 j_{i} (t, x_{i}) := - \frac{i\hbar}{2m} 
  \left\{
    \Phi_{i}^{\ast} (t, x_{i}) \Phi'_{i} (t, x_{i}) 
    - {\Phi_{i}^{\ast}}' (t, x_{i}) \Phi_{i} (t, x_{i}) 
  \right\} ,
\end{equation}
and we implicitly assume the limit
\begin{equation}
 j_{i} (t, \xi) := \lim_{x_{i} \rightarrow \xi} j_{i}(t, x_{i}) .
\end{equation}  
Here the prime $(')$ denotes the partial differentiation 
with respect to the spatial coordinate.
Equation (\ref{eq:c-conserv}) can be expressed by 
\begin{equation}
\label{eq:jc1}
 \Psi'^{\dagger} \Psi - \Psi^{\dagger} \Psi' = 0 ,
\end{equation}
where 
\begin{equation}
 \Psi : = \left( 
  \begin{array}{c}
   \Phi_{1} (t, \xi ) \\ \Phi_{2} (t, \xi ) \\ \Phi_{3} (t, \xi ) 
  \end{array}
 \right) , \quad 
 \Psi' : = \left( 
  \begin{array}{c}
   \Phi'_{1} (t, \xi ) \\ \Phi'_{2} (t, \xi ) \\ \Phi'_{3} (t, \xi ) 
  \end{array}
 \right) .
\end{equation} 
Equation (\ref{eq:jc1}) is equivalently expressed as \cite{cft}
\begin{equation}
 \left| \Psi - i L_{0} \Psi'\right|
 = \left| \Psi + i L_{0} \Psi'\right| , 
\end{equation}
where $L_{0}$ $(\in \mathbb{R} )$ is an arbitrary
non-vanishing constant with dimension of length
(see \ref{sec:a-l0} for the role of $L_{0}$).
Hence $\Psi - i L_{0} \Psi'$ is connected 
to $\Psi + i L_{0} \Psi'$ via a unitary transformation. 
Thus, we obtain the junction condition \cite{cft}
\begin{equation}
\label{eq:jc2}
 \left( U-I_{3} \right) \Psi 
  + i L_{0} \left( U+I_{3}\right) \Psi' = 0 ,
\end{equation}
where $I_{3}$ is the $3\times 3$ identity matrix, 
and $U$ is a $3\times 3 $ unitary matrix, i.e., $U \in {\rm U}(3)$.
Therefore the junction condition is characterized by 
the unitary matrix $U$.

Next, we discuss parametrization of the unitary matrix $U$.
For this discussion, it is useful to recall that 
any unitary matrix $U$ can be diagonalized by 
a unitary matrix $W$ as
\begin{equation}
\label{eq:diag-w}
 W^{\dagger} U W = D 
 :=\left(
 \begin{array}{ccc}
  e^{i\theta_{1}} & 0 & 0 \\
  0 & e^{i\theta_{2}} & 0 \\
  0 &  0 & e^{i\theta_{3}}  \\
 \end{array}
 \right) 
\end{equation}
where $\theta_{i} \in \mathbb{R} \ (i=1,2,3)$.
Since ${\rm U}(3) = {\rm U}(1) \times {\rm SU}(3)$
holds locally \cite{zwiebach},  
the unitary matrix $W$ can
always be expressed by an element $V \in$ SU(3)
multiplied by a complex factor $e^{i\eta}$,  i.e.,
$W=e^{i\eta} V$,  where $\eta \in \mathbb{R}$. 
When we adopt the Euler angle parametrization \cite{eap}
for $V$, we have
\begin{equation}
 W=e^{i\eta} e^{i\alpha \lambda_{3}} e^{i\beta \lambda_{2}}
   e^{i\gamma \lambda_{3}} e^{i\delta \lambda_{5}}
   e^{i a \lambda_{3}} e^{ib \lambda_{2}}
   e^{i c \lambda_{3}} e^{id \lambda_{8}}
\end{equation}
where $\lambda_{1}, \lambda_{2}, \cdots , \lambda_{8}$ 
are the Gell-Mann Matrices 
(see \ref{sec:a-gm} for the definition), 
and $\alpha, \beta , \gamma , \delta , a, b, c, d
\in \mathbb{R}$.
Then we derive
\begin{equation}
\label{eq:u-exp}
 U=WDW^{\dagger}
   ={\cal V} D {\cal V}^{\dagger},
\end{equation}
where
\begin{equation}
\label{eq:V}
 {\cal V}= e^{i\alpha \lambda_{3}} e^{i\beta \lambda_{2}}
   e^{i\gamma \lambda_{3}} e^{i\delta \lambda_{5}}
   e^{i a \lambda_{3}} e^{ib \lambda_{2}} .
\end{equation}
Thus the unitary matrix $U$, which provides the junction
condition, is specified by 
the nine real parameters: $\theta_{1}$, $\theta_{2}$,
$\theta_{3}$, $\alpha$, $\beta$, $\gamma$,
$\delta$, $a$ and $b$.

\subsection{The scattering matrices}

Let us consider the scattering of a single mode
with wave number $k$ by a Y-junction.
We assume that incoming waves and outgoing waves
are provided by $\phi_{i} e^{ikx_{i}}$ 
and $\psi_{i} e^{-ikx_{i}}$ $(i=1,2,3)$, respectively,
as shown in Fig.~\ref{fig1}(a). 
Here $\phi_{i} , \psi_{i} \in \mathbb{C}$.
Then, we have 
\begin{equation}
 \Phi_{i} \left( t, x_{i}\right) 
 = e^{-i\frac{\cal E}{\hbar}t} 
 \left( \phi_{i} e^{ik x_{i}} + \psi_{i} e^{-ikx_{i}} \right) ,
\end{equation}
and
\begin{equation}
\label{eq:Psi}
 \Psi = e^{-i \frac{\cal E}{\hbar} t} \left\{ e^{ik \xi}
  \left( 
  \begin{array}{c}
   \phi_{1}  \\  \phi_{2}  \\  \phi_{3} 
  \end{array}
 \right)  + e^{-ik \xi} 
 \left( 
  \begin{array}{c}
   \psi_{1}  \\  \psi_{2}  \\  \psi_{3} 
  \end{array}
 \right) 
 \right\}, 
\end{equation}
\begin{equation}
\label{eq:Psid}
 \Psi' = e^{-i\frac{\cal E}{\hbar}t} \left\{ ik e^{ik \xi}
  \left( 
  \begin{array}{c}
   \phi_{1}  \\  \phi_{2}  \\  \phi_{3} 
  \end{array}
 \right)  -ik e^{-ik \xi} 
 \left( 
  \begin{array}{c}
   \psi_{1}  \\  \psi_{2}  \\  \psi_{3} 
  \end{array}
 \right) 
 \right\}, 
\end{equation}
where ${\cal E}:=\hbar^2 k^2/2m$. 
Substituting Eqs.~(\ref{eq:Psi}) and (\ref{eq:Psid}) into 
Eq.~(\ref{eq:jc2}) and using Eq.~(\ref{eq:u-exp}), 
we can rewrite the expression (\ref{eq:jc2}) into the form
\begin{equation}
\label{eq:in-to-out}
  \left( 
  \begin{array}{c}
   \psi_{1}  \\  \psi_{2}  \\  \psi_{3} 
  \end{array}
 \right) =S (\xi)
 \left( 
  \begin{array}{c}
   \phi_{1}  \\  \phi_{2}  \\  \phi_{3} 
  \end{array}
 \right) .
\end{equation}
Based on the inward axes as shown in Fig.~\ref{fig1}(a), 
we derive the $S$-matrix as
\begin{equation}
\label{eq:s-matrix}
 S (\xi )= S^{\rm (in)} (\xi )
 :=e^{2ik\xi} {\cal V} S_{0}^{\rm (in)} {\cal V}^{\dagger} ,
\end{equation}
where 
\begin{equation}
 S_{0}^{\rm (in)} :=\left( 
 \begin{array}{ccc}
  \frac{ikL_{1}+1}{ikL_{1}-1} & 0 & 0\\
  0 & \frac{ikL_{2}+1}{ikL_{2}-1} & 0 \\
  0 & 0 & \frac{ikL_{3}+1}{ikL_{3}-1} \\
 \end{array}
 \right) ,
\end{equation}
and
\begin{equation}
\label{eq:li}
 L_{i} := L_{0} \cot\frac{\theta_{i}}{2} .
\end{equation}
Here the superscript ``(in)'' denotes the result derived 
on the inward axes.
Note that the diagonal component $S_{ii}$ $(i=1,2,3)$ represents 
the probability amplitude for the reflection 
from $x_{i}$-axis to $x_{i}$-axis, 
while the non-diagonal component $S_{ij}$ ($i\neq j$) 
represents the probability amplitude 
for the transmission from $x_{j}$-axis to $x_{i}$-axis. 
The $S$-matrix  (\ref{eq:s-matrix}) is not symmetric in general, 
because the junction conditions do not necessarily satisfy the time-reversal symmetry
(see \ref{sec:a-tr} for the condition of the time-reversal symmetry).

It should also be noted that when we take the axes in the
opposite direction as shown in Fig.~\ref{fig1}(b) (outward axes),
we have to replace the wave number $k$ with $-k$.
In this case, we derive the $S$-matrix in the form
\begin{equation}
 S(\xi ) = S^{\rm (out)} (\xi )
 :=e^{-2ik\xi} {\cal V} S_{0}^{\rm (out)} {\cal V}^{\dagger} ,
\end{equation}
\begin{equation}
 S_{0}^{\rm (out)}:=\left( 
 \begin{array}{ccc}
  \frac{ikL_{1}-1}{ikL_{1}+1} & 0 & 0\\
  0 & \frac{ikL_{2}-1}{ikL_{2}+1} & 0 \\
  0 & 0 & \frac{ikL_{3}-1}{ikL_{3}+1} \\
 \end{array}
 \right) .
\end{equation}
Here the superscript ``(out)'' denotes the result derived 
on the outward axes.
This expression is also useful to discuss 
a ring system with double Y-junctions in the next section.

\section{Formulations of a ring system with double Y-junctions}
\label{sec3}

\subsection{Preliminary}
\label{sec3-1}

\begin{figure}[tb]
  \includegraphics[width=.45 \textwidth]{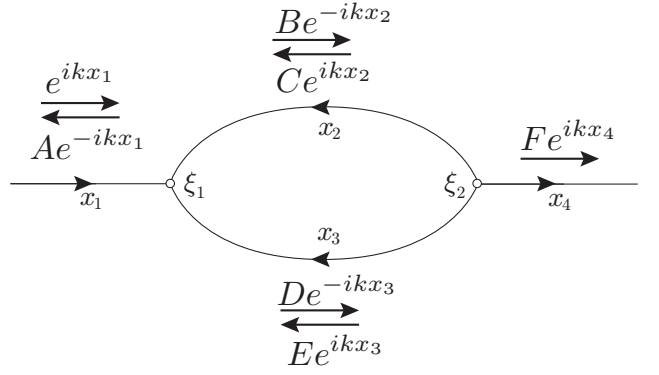}
  \caption{A ring system with double Y-junctions.  
          }
  \label{fig2}
\end{figure}

We consider a ring system made of double Y-junctions
as shown in Fig.~\ref{fig2}. In this system,
the Y-junction on the left has the inward axes of
$x_{1}$, $x_{2}$ and $x_{3}$, while the 
 Y-junction on the right has the outward axes
of $x_{2}$, $x_{3}$ and $x_{4}$.
The nodes are given by 
$x_{1}=x_{2}=x_{3}=\xi_{1}$ 
and $x_{2}=x_{3}=x_{4}=\xi_{2}$, where
$\xi_{1} > \xi_{2}$. 
We also assume that 
the node on the Y-junction $(x_{1}x_{2}x_{3})$
has the same parameters as the node
on the Y-junction $(x_{4}x_{2}x_{3})$ symmetrically,
as shown in Fig.~\ref{fig3}(a).
We call this type of a ring {\it a symmetric ring}.
(We also discuss {\it an anti-symmetric ring} below.)
When we consider a scattering problem, 
we take the wave functions on the wires as 
\begin{eqnarray}
\label{eq:yj-phi1}
 \Phi_{1} (t,x_{1})& =  & 
   e^{-i\frac{\cal E}{\hbar}t} \left( e^{ikx_{1}} + Ae^{-ikx_{1}} \right) , \\
\label{eq:yj-phi2}
 \Phi_{2} (t,x_{2})& =  & 
   e^{-i\frac{\cal E}{\hbar}t} \left( B e^{-ikx_{2}} + C e^{ikx_{2}} \right) , \\
\label{eq:yj-phi3}
 \Phi_{3} (t,x_{3})& =  & 
   e^{-i\frac{\cal E}{\hbar}t} \left( D e^{-ikx_{3}} + E e^{ikx_{3}} \right) , \\
\label{eq:yj-phi4}
 \Phi_{4} (t,x_{4})& =  & 
   e^{-i\frac{\cal E}{\hbar}t} Fe^{ikx_{4}} ,
\end{eqnarray}
where $A, B, C, D, E, F \in \mathbb{C}$.
Applying these expressions to Eq.~(\ref{eq:in-to-out}),
we have the relations
\begin{equation}
\label{eq:c-r1}
  \left( 
  \begin{array}{c}
   A  \\  B  \\  D 
  \end{array}
 \right) = S_{1}^{\rm (in)} (\xi_{1} )
 \left( 
  \begin{array}{c}
   1  \\  C  \\  E
  \end{array}
 \right) , 
\end{equation}
\begin{equation}
\label{eq:c-r2}
  \left( 
  \begin{array}{c}
  F \\  C  \\  E   
  \end{array}
 \right) = S_{2}^{\rm (out)} (\xi_{2} )
 \left( 
  \begin{array}{c}
   0 \\ B  \\  D  
  \end{array}
 \right) ,
\end{equation}
where $S_{1}$ denotes the $S$-matrix
on the Y-junction $(x_{1}x_{2}x_{3})$, and 
$S_{2}$ denotes the $S$-matrix
on the Y-junction $(x_{4}x_{2}x_{3})$. 
By solving Eqs.~(\ref{eq:c-r1}) and (\ref{eq:c-r2}),  
we can derive the coefficients $A, B, C, D, E$ and $F$
as functions of the components of $S_{1}^{\rm (in)} (\xi_{1} )$
and $S_{2}^{\rm (out)} (\xi_{2} )$ (see \ref{sec:s-a}).

\begin{figure}[tb]
  \includegraphics[width=.4 \textwidth]{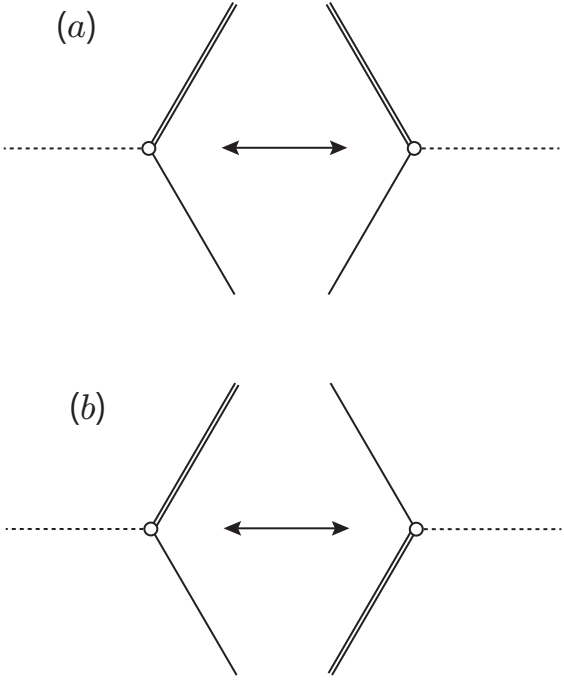}
  \caption{(a) A symmetric ring system. 
         (b) An anti-symmetric ring system.  
          }
  \label{fig3}
\end{figure}

\subsection{Path integral description}

Let us derive the solution for the coefficients $A, B, C, D, E$ 
and $F$ in Eqs.~(\ref{eq:yj-phi1})-(\ref{eq:yj-phi4})
based on a path integral approach.
Our approach is similar to the method of a path decomposition 
expansion developed in \cite{ak}
(see \cite{ohya, os,ohya2} for general treatments).
The path integral makes physical interpretation clear.

First, we obtain the solution for $F$.
We now consider all possible paths from $x_{1}$ to $x_{4}$. 
Each path can be characterized by the number of 
scattering at the nodes. Let us assume that 
$F_{n}$ denotes the contribution from the path 
having $n$ times scattering to $F$. Here
$n=2\ell_{1} \ (\ell_{1}=1,2,3, \cdots)$ is only permitted.
Thus, we have 
\begin{equation}
 F=F_{2}+F_{4}+F_{6}+\cdots .
\end{equation}
For example, $F_{2}$ is composed of two paths:
($x_{1} \rightarrow x_{2} \rightarrow x_{4}$) and 
($x_{1} \rightarrow x_{3} \rightarrow x_{4}$). 
Hence, when we write the components of $S^{\rm (in)}(\xi_{1})$
and $S^{\rm (out)}(\xi_{2})$ as
\begin{equation}
 S^{\rm (in)}_{1} (\xi_{1} ) := \left( 
 \begin{array}{ccc}
  s_{11} & s_{12} & s_{13} \\
  s_{21} & s_{22} & s_{23} \\
  s_{31} & s_{32} & s_{33} \\
 \end{array}
 \right) ,
\end{equation}
\begin{equation}
 S^{\rm (out)}_{2} (\xi_{2} )= \left( 
 \begin{array}{ccc}
  \tilde{s}_{11} & \tilde{s}_{12} & \tilde{s}_{13} \\
  \tilde{s}_{21} & \tilde{s}_{22} & \tilde{s}_{23} \\
  \tilde{s}_{31} & \tilde{s}_{32} & \tilde{s}_{33} \\
 \end{array}
 \right) ,
\end{equation}
we have
\begin{equation}
 F_{2}= \tilde{s}_{12} s_{21} + \tilde{s}_{13} s_{31} 
    =\sum_{i=2,3} \tilde{s}_{1i} s_{i1} .
\end{equation}
Similarly, we derive
\begin{eqnarray}
 F_{4} & = & \sum_{i_{n}=2,3} 
   \tilde{s}_{1i_{1}} s_{i_{1}i_{2}} 
   \tilde{s}_{i_{2}i_{3}} s_{i_{3}1} , \\
 F_{6} & = & \sum_{i_{n}=2,3} 
   \tilde{s}_{1i_{1}} s_{i_{1}i_{2}} \tilde{s}_{i_{2}i_{3}} 
   s_{i_{3}i_{4}} \tilde{s}_{i_{4}i_{5}} s_{i_{5}1} , \\
 & \vdots &  \nonumber \\
 F_{2\ell_{1}} & = & \sum_{i_{n}=2,3} 
   \tilde{s}_{1i_{1}} s_{i_{1}i_{2}} \tilde{s}_{i_{2}i_{3}} 
   s_{i_{3}i_{4}} \tilde{s}_{i_{4}i_{5}} 
   \cdots s_{i_{2\ell_{1}-1}1} .
\end{eqnarray}
When we define $2\times 2$ matrices
\begin{equation}
 s:= \left( 
 \begin{array}{cc}
  s_{22} & s_{23} \\ s_{32} & s_{33} 
 \end{array}
 \right) ,
\end{equation}
\begin{equation}
 \tilde{s} := \left( 
 \begin{array}{cc}
  \tilde{s}_{22} & \tilde{s}_{23} \\ 
  \tilde{s}_{32} & \tilde{s}_{33} 
 \end{array}
 \right) ,
\end{equation}
then we derive
\begin{eqnarray}
 F_{2} & = & 
 \left( 
 \begin{array}{cc}
  \tilde{s}_{12} & \tilde{s}_{13}\\
 \end{array}
 \right) \left( 
 \begin{array}{c}
  s_{21} \\ s_{31}
 \end{array}
 \right) , \\
 F_{4} & = & 
 \left( 
 \begin{array}{cc}
  \tilde{s}_{12} & \tilde{s}_{13}\\
 \end{array}
 \right) s\tilde{s} \left( 
 \begin{array}{c}
  s_{21} \\ s_{31}
 \end{array}
 \right) , \\
 F_{6} & = & 
 \left( 
 \begin{array}{cc}
  \tilde{s}_{12} & \tilde{s}_{13}\\
 \end{array}
 \right) \left( s\tilde{s} \right)^2 \left( 
 \begin{array}{c}
  s_{21} \\ s_{31}
 \end{array}
 \right) , \\
 & \vdots & \nonumber \\
  F_{2\ell_{1}} & = & 
 \left( 
 \begin{array}{cc}
  \tilde{s}_{12} & \tilde{s}_{13}\\
 \end{array}
 \right) \left( s\tilde{s} \right)^{\ell_{1}-1} \left( 
 \begin{array}{c}
  s_{21} \\ s_{31}
 \end{array}
 \right) .
\end{eqnarray}
Therefore, we obtain the amplitude $F$ as
\begin{equation}
\label{eq:F0}
 F =  
 \left( 
 \begin{array}{cc}
  \tilde{s}_{12} & \tilde{s}_{13}\\
 \end{array}
 \right) \left\{
 I_{2} + s\tilde{s} + \left( s\tilde{s} \right)^2 
 + \cdots 
 \right\} \left( 
 \begin{array}{c}
  s_{21} \\ s_{31}
 \end{array}
 \right) , 
\end{equation}
where $I_{2}$ is the $2\times 2$ identity matrix.

Next, we obtain the solution for $A$. 
We consider all possible paths from $x_{1}$
to $x_{1}$. Let us assume that 
$A_{n}$ denotes the contribution from the path 
having $n$ times scattering to $A$. 
Here $n=2\ell_{2}+1 \ (\ell_{2}=0,1,2, \cdots )$.
Then we have
\begin{equation}
 A=A_{1} + A_{3} + A_{5} +\cdots .
\end{equation}
For example, $A_{1}$ is composed of the one path: 
($x_{1} \rightarrow x_{1}$), and 
$A_{3}$ is composed of four paths:
($x_{1} \rightarrow x_{2} \rightarrow x_{2} \rightarrow x_{1}$),
($x_{1} \rightarrow x_{2} \rightarrow x_{3} \rightarrow x_{1}$), 
($x_{1} \rightarrow x_{3} \rightarrow x_{2} \rightarrow x_{1}$) and 
($x_{1} \rightarrow x_{3} \rightarrow x_{3} \rightarrow x_{1}$).  
Thus we derive
\begin{eqnarray}
 A_{1} & = & s_{11} \\
 A_{3} & = & 
 \left( 
 \begin{array}{cc}
  s_{12} & s_{13}\\
 \end{array}
 \right) \tilde{s}\left( 
 \begin{array}{c}
  s_{21} \\ s_{31}
 \end{array}
 \right) , \\
 A_{5} & = & 
 \left( 
 \begin{array}{cc}
  s_{12} & s_{13}\\
 \end{array}
 \right) \tilde{s}s\tilde{s} \left( 
 \begin{array}{c}
  s_{21} \\ s_{31}
 \end{array}
 \right) , \\
 & \vdots & \nonumber \\
 A_{2\ell_{2}+1} & = & 
 \left( 
 \begin{array}{cc}
  s_{12} & s_{13}\\
 \end{array}
 \right)  \tilde{s} \left( s\tilde{s} \right)^{\ell_{2}-1}\left( 
 \begin{array}{c}
  s_{21} \\ s_{31}
 \end{array}
 \right)  \ \  (\ell_{2} \geq 1).
\end{eqnarray}
Therefore we obtain
\begin{equation}
\label{eq:A0}
 A= s_{11} +  \left( 
 \begin{array}{cc}
  s_{12} & s_{13}\\
 \end{array}
 \right) \tilde{s} \left\{ 
   I_{2} + s\tilde{s} + \left( s\tilde{s} \right)^2
   +\cdots \right\} \left( 
 \begin{array}{c}
  s_{21} \\ s_{31}
 \end{array}
 \right) .
\end{equation}
The coefficient $A$ is expressed by the sum of 
a series of $s\tilde{s}$ in the same manner as $F$.

Finally, in a similar way, we obtain the other coefficients 
$B, C, D$ and $E$ as
\begin{equation}
\label{eq:B0}
 B= s_{21} +  \left( 
 \begin{array}{cc}
  s_{22} & s_{23}\\
 \end{array}
 \right) \tilde{s} \left\{ 
   I_{2} + s\tilde{s} + \left( s\tilde{s} \right)^2
   +\cdots \right\} \left( 
 \begin{array}{c}
  s_{21} \\ s_{31}
 \end{array}
 \right) ,
\end{equation}
\begin{equation}
\label{eq:C0}
 C= \left( 
 \begin{array}{cc}
  \tilde{s}_{22} & \tilde{s}_{23}\\
 \end{array}
 \right) \left\{ 
   I_{2} + s\tilde{s} + \left( s\tilde{s} \right)^2
   +\cdots \right\} \left( 
 \begin{array}{c}
  s_{21} \\ s_{31}
 \end{array}
 \right) ,
\end{equation}
\begin{equation}
\label{eq:D0}
 D= s_{31} +  \left( 
 \begin{array}{cc}
  s_{32} & s_{33}\\
 \end{array}
 \right) \tilde{s} \left\{ 
   I_{2} + s\tilde{s} + \left( s\tilde{s} \right)^2
   +\cdots \right\} \left( 
 \begin{array}{c}
  s_{21} \\ s_{31}
 \end{array}
 \right) ,
\end{equation}
\begin{equation}
\label{eq:E0}
 E= \left( 
 \begin{array}{cc}
  \tilde{s}_{32} & \tilde{s}_{33}\\
 \end{array}
 \right)  \left\{ 
   I_{2} + s\tilde{s} + \left( s\tilde{s} \right)^2
   +\cdots \right\} \left( 
 \begin{array}{c}
  s_{21} \\ s_{31}
 \end{array}
 \right) .
\end{equation}
Thus the coefficients $B, C, D$ and $E$ are also 
expressed by the sum of a series of $s\tilde{s}$.

The above expressions can be simplified.
Since $\left| s_{ij} \right| <1$ and  
$\left| \tilde{s}_{ij} \right| <1$ ($i,j=2,3$), 
we have
\begin{equation}
 \lim_{n\rightarrow \infty}
   \left( s\tilde{s} \right)^{n} =O_2 , 
\end{equation}
where $O_{2}$ is the $2\times 2$ zero matrix.
Hence the series of $s\tilde{s}$ converges as
\begin{equation}
 I_{2} + s\tilde{s} + \left( s\tilde{s} \right)^2
   +\cdots
 = \left( I_{2} - \left( s\tilde{s} \right) \right)^{-1} .
\end{equation}
Therefore we can rewrite the above results as
\begin{equation}
\label{eq:A}
 A= s_{11} +  \left( 
 \begin{array}{cc}
  s_{12} & s_{13}\\
 \end{array}
 \right) \tilde{s}
 \left( I_{2} - \left( s\tilde{s} \right) \right)^{-1} \left( 
 \begin{array}{c}
  s_{21} \\ s_{31}
 \end{array}
 \right) .
\end{equation}
\begin{equation}
 B= s_{21} +  \left( 
 \begin{array}{cc}
  s_{22} & s_{23}\\
 \end{array}
 \right) \tilde{s} 
 \left( I_{2} - \left( s\tilde{s} \right) \right)^{-1} 
 \left( 
 \begin{array}{c}
  s_{21} \\ s_{31}
 \end{array}
 \right) ,
\end{equation}
\begin{equation}
 C= \left( 
 \begin{array}{cc}
  \tilde{s}_{22} & \tilde{s}_{23}\\
 \end{array}
 \right) 
 \left( I_{2} - \left( s\tilde{s} \right) \right)^{-1}
 \left( 
 \begin{array}{c}
  s_{21} \\ s_{31}
 \end{array}
 \right) ,
\end{equation}
\begin{equation}
 D= s_{31} +  \left( 
 \begin{array}{cc}
  s_{32} & s_{33}\\
 \end{array}
 \right) \tilde{s} 
 \left( I_{2} - \left( s\tilde{s} \right) \right)^{-1}
 \left( 
 \begin{array}{c}
  s_{21} \\ s_{31}
 \end{array}
 \right) ,
\end{equation}
\begin{equation}
 E= \left( 
 \begin{array}{cc}
  \tilde{s}_{32} & \tilde{s}_{33}\\
 \end{array}
 \right) 
   \left( I_{2} - \left( s\tilde{s} \right) \right)^{-1}
   \left( 
 \begin{array}{c}
  s_{21} \\ s_{31}
 \end{array}
 \right) ,
\end{equation}
\begin{equation}
\label{eq:F}
 F =  
 \left( 
 \begin{array}{cc}
  \tilde{s}_{12} & \tilde{s}_{13}\\
 \end{array}
 \right)
 \left( I_{2} - \left( s\tilde{s} \right) \right)^{-1}
 \left( 
 \begin{array}{c}
  s_{21} \\ s_{31}
 \end{array}
 \right) .
\end{equation}
By straightforward calculations, we can confirm 
that the above results coincide with the results
derived by solving Eqs.~(\ref{eq:c-r1}) 
and (\ref{eq:c-r2}) algebraically (see \ref{sec:s-a}). 
Using Eqs.~(\ref{eq:A}) and (\ref{eq:F}), we 
can calculate the probability for reflection $|A|^2$
and that for transmission $|F|^2$.

\section{Systems under scale-invariant junction conditions}
\label{sec4}

In this section, we restrict our attention to the cases of 
scale-invariant junction conditions for explicit examples.

\subsection{A system with a single Y-junction}

We consider a system with a single Y-junction.
Under the scale-invariant junction conditions, the eigen values of $U$ 
take $+1$ or $-1$, that is, $\theta_{i}$ takes $0$ or $\pi$
(see \ref{sec:apd-sij} for the details).
Then, the diagonal matrix $D$ is given by 
\begin{equation}
\label{eq:ds}
 D=D_{S}:= \left( 
 \begin{array}{ccc}
 \epsilon (\theta_{1}) & 0 & 0 \\
 0 & \epsilon (\theta_{2}) & 0 \\
 0 & 0 & \epsilon (\theta_{3})  \\
 \end{array}
 \right) ,
\end{equation}
where
\begin{equation}
 \epsilon (\theta_{i})
 = \left\{
 \begin{array}{lc}
  +1 & (\theta_{i}=0 )\\
  -1 & (\theta_{i}=\pi )\\
 \end{array}
 \right. .
\end{equation}
Substituting Eqs.~(\ref{eq:u-exp}) and (\ref{eq:ds}) 
into Eq.~(\ref{eq:jc2}), we find that the junction 
condition is divided into two conditions
\begin{equation}
 \left( U-I_{3} \right) \Psi = 0,
\end{equation}
\begin{equation}
 \left( U+I_{3} \right) \Psi' = 0.
\end{equation}
Thus the dimensional parameter $L_{0}$ is dropped
in the junction conditions.
Furthermore, from Eq.~(\ref{eq:li}),  we have 
\begin{equation}
 L_{i} \rightarrow \pm\infty \quad \mbox{as} \quad \theta_{i}=0 ,
\end{equation}
\begin{equation}
 L_{i} =0 \quad \mbox{as} \quad \theta_{i}=\pi .
\end{equation}
Hence we derive 
\begin{equation}
 S_{0}^{\rm (in)} = S_{0}^{\rm (out)} = D_{S} .
\end{equation}
It should be noted that the $k$-dependence of
$S_{0}^{\rm (in)}$ and $S_{0}^{\rm (out)}$ disappears
under the scale-invariant junction conditions.
Thus, we obtain the $S$-matrix
\begin{eqnarray}
\label{eq:si-sm1}
  S^{\rm (in)} (\xi ) 
    & = & e^{2ik\xi} {\cal V} D_{S} {\cal V}^{\dagger} ,\\
\label{eq:si-sm2}
  S^{\rm (out)} (\xi ) 
    & = & e^{-2ik\xi} {\cal V} D_{S} {\cal V}^{\dagger} .
\end{eqnarray}
Since the three parameters $\theta_{i} \ (i=1,2,3)$ are fixed, 
six parameters remain in the $S$-matrices. 
From Eqs.~(\ref{eq:si-sm1}) and (\ref{eq:si-sm2}), we find the 
probability for reflection from the $x_{i}$-axis 
to the $x_{i}$-axis
\begin{equation}
 P\left( i \rightarrow i \right)
 = \left| \left( {\cal V} D_{S} {\cal V}^{\dagger} \right)_{ii} 
 \right|^2 ,
\end{equation}
and the probability for transmission 
from the $x_{i}$-axis to the $x_{j}$-axis
\begin{equation}
 P\left( i \rightarrow j \right)
 = \left| \left( {\cal V} D_{S} {\cal V}^{\dagger} \right)_{ji} 
 \right|^2 .
\end{equation}
In these expressions, $k$ does not appear.
Therefore, when we consider the scale-invariant 
junction conditions, the probabilities for reflection 
and transmission become constant with respect to $k$.

\subsection{A symmetric ring system with double Y-junctions}

We consider a symmetric ring system with double Y-junctions
described in Sec.~\ref{sec3-1} (see also Fig.~\ref{fig3}(a)). 
From Eqs.~(\ref{eq:si-sm1}) and 
(\ref{eq:si-sm2}), we have
the $S$ matrix at $\xi_{1}$ and that at $\xi_{2}$ as
\begin{equation}
 S_{1}^{\rm (in)} (\xi_{1} )=
   e^{2ik\xi_{1}} {\cal V} D_{S} {\cal V}^{\dagger} ,
\end{equation}
\begin{equation}
 S_{2}^{\rm (out)} (\xi_{2} )=
   e^{-2ik\xi_{2}} {\cal V} D_{S} {\cal V}^{\dagger} .
\end{equation}
We drop the superscripts ``(in)" and ``(out)" in what follows.
From these equations, we find the relations
\begin{equation}
\label{eq:ss1}
 S_{1} (\xi_{1} ) S_{1} (\xi_{1} )^{\dagger}
  = S_{1} (\xi_{1} )^{\dagger}  S_{1} (\xi_{1} )  = I_{3} ,
\end{equation}
\begin{equation}
\label{eq:ss2}
 S_{2} (\xi_{2} ) S_{2} (\xi_{2} )^{\dagger}
  = S_{2} (\xi_{2} )^{\dagger}  S_{2} (\xi_{2} )  = I_{3} ,
\end{equation}
\begin{equation}
\label{eq:ss2}
 S_{1} (\xi_{1} ) S_{1} (\xi_{1} )  = e^{4ik\xi_{1}} ,
\end{equation}
\begin{equation}
\label{eq:ss4}
 S_{2} (\xi_{2} ) S_{2} (\xi_{2} )  = e^{-4ik\xi_{2}} ,
\end{equation}
\begin{equation}
\label{eq:ss5}
 S_{2} (\xi_{2} )=
   e^{2ik(\xi_{1} - \xi_{2})} S_{1} (\xi_{1} )^{\dagger} .
\end{equation}
From Eqs.~(\ref{eq:ss1}) and (\ref{eq:ss5}), 
we can derive
\begin{equation}
 \tilde{s} =  e^{2ik(\xi_{1} - \xi_{2})} 
 \left( 
 \begin{array}{cc}
  s_{22}^{\ast}  &  s_{32}^{\ast}\\
  s_{23}^{\ast}  &  s_{33}^{\ast}
 \end{array}
 \right) ,
\end{equation}
and
\begin{equation}
 s\tilde{s} =  e^{2ik(\xi_{1} - \xi_{2})} 
 \left( 
 \begin{array}{cc}
  \left| s_{22} \right|^2 +\left| s_{23} \right|^2 
   & -s_{21} s_{31}^{\ast}\\
  -s_{21}^{\ast} s_{31} 
  & \left| s_{32} \right|^2 +\left| s_{33} \right|^2 
 \end{array}
 \right) .
\end{equation}
Furthermore, we obtain
\begin{equation}
\label{eq:ss_eigen}
 s\tilde{s} 
 \left( 
 \begin{array}{c}
  s_{21} \\ s_{31}
 \end{array}
 \right) = e^{2ik(\xi_{1} - \xi_{2})} 
 \left| s_{11} \right|^2 
 \left( 
 \begin{array}{c}
  s_{21} \\ s_{31}
 \end{array}
 \right) .
\end{equation}
Thus an eigen value and an eigen vector 
of $s\tilde{s}$ are given 
by $e^{2ik(\xi_{1} - \xi_{2})}  \left| s_{11} \right|^2 $
and $\displaystyle \left( \begin{array}{c}
  s_{21} \\ s_{31} \end{array} \right)$, respectively.
Substituting Eq.~(\ref{eq:ss_eigen}) into 
Eqs.~(\ref{eq:A0})-(\ref{eq:E0}) and (\ref{eq:F0}),
we derive the simplified results
\begin{eqnarray}
\label{eq:si-A}
 A & = & \frac{(1-e^{2ik(\xi_{1} - \xi_{2})}) s_{11}}
  {1-e^{2ik(\xi_{1} - \xi_{2})} \left| s_{11}\right|^2} , \\
\label{eq:si-B}
 B & = & \frac{s_{21}}
  {1-e^{2ik(\xi_{1} - \xi_{2})} \left| s_{11}\right|^2} , \\
\label{eq:si-C}
 C & = & \frac{-e^{2ik(\xi_{1} - \xi_{2})} s_{11} s_{12}^{\ast}}
  {1-e^{2ik(\xi_{1} - \xi_{2})} \left| s_{11}\right|^2} , \\
\label{eq:si-D}
 D & = & \frac{s_{31}}
  {1-e^{2ik(\xi_{1} - \xi_{2})} \left| s_{11}\right|^2} , \\
\label{eq:si-E}
 E & = & \frac{-e^{2ik(\xi_{1} - \xi_{2})} s_{11} s_{13}^{\ast}}
  {1-e^{2ik(\xi_{1} - \xi_{2})} \left| s_{11}\right|^2} , \\
\label{eq:si-F}
 F & = & \frac{e^{2ik(\xi_{1} - \xi_{2})} 
  \left( 1- \left|s_{11} \right|^2\right)}
  {1-e^{2ik(\xi_{1} - \xi_{2})} \left| s_{11}\right|^2} .
\end{eqnarray}
Consequently, from Eqs.~(\ref{eq:si-A}) and (\ref{eq:si-F}), 
we find that the perfect transmission ($A=0$)
occurs if and only if the condition 
\begin{equation}
 e^{2ik (\xi_{1} - \xi_{2})} = 1
\end{equation}
holds except the trivial case of  $s_{11} = 0$.
On the other hand, the perfect reflection ($F=0$)
does not occur except the trivial case of $|s_{11}| = 1$.

\subsection{An anti-symmetric ring system with double Y-junctions}

We also discuss an anti-symmetric ring system 
in which the $x_{2}$-components in the $S$-matrix
are replaced with the $x_{3}$-components 
and vice versa on the Y-junction $(x_{4}x_{2}x_{3})$ 
on the right, as shown in Fig.~\ref{fig3}(b).
That is, we replace
$S_{2} (\xi_{2} )$ with $\overline{S}_{2} (\xi_{2} )$, where
\begin{equation}
   \overline{S}_{2} (\xi_{2} )  := P S_{2} (\xi_{2} ) P^{-1} ,
\end{equation} 
Here $P$ is defined by  
\begin{equation}
 P := \left(
  \begin{array}{ccc}
   1 & 0 & 0 \\ 0 & 0 & 1 \\ 0 & 1 & 0 \\
  \end{array} 
 \right) .
\end{equation}
Under the scale-invariant junction conditions, 
from Eq.~(\ref{eq:ss5}), we derive
\begin{eqnarray}
 \overline{S}_{2} (\xi_{2} )
 & = & 
   e^{2ik(\xi_{1} - \xi_{2})} P S_{1} (\xi_{1} )^{\dagger} P^{-1}  
   \nonumber \\
 & = &  e^{2ik(\xi_{1} - \xi_{2})}
  \left(
  \begin{array}{ccc}
   s_{11}^{\ast} & s_{31}^{\ast} & s_{21}^{\ast} \\ 
   s_{13}^{\ast} & s_{33}^{\ast} & s_{23}^{\ast} \\ 
   s_{12}^{\ast} & s_{32}^{\ast} & s_{22}^{\ast} 
  \end{array} 
 \right) .
\end{eqnarray}
Thus, we have
\begin{equation}
 \tilde{s} =  e^{2ik(\xi_{1} - \xi_{2})} 
 \left( 
 \begin{array}{cc}
  s_{33}^{\ast}  &  s_{23}^{\ast}\\
  s_{32}^{\ast}  &  s_{22}^{\ast}
 \end{array}
 \right) ,
\end{equation}
and
\begin{equation}
 s\tilde{s} =  e^{2ik(\xi_{1} - \xi_{2})} 
 \left( 
 \begin{array}{cc}
  s_{22} s_{33}^{\ast} +s_{23} s_{32}^{\ast} 
   & s_{22} s_{23}^{\ast}+s_{23} s_{22}^{\ast}\\
  s_{32} s_{33}^{\ast} +s_{33} s_{32}^{\ast}
  & s_{32} s_{23}^{\ast} +s_{33} s_{22}^{\ast}
 \end{array}
 \right) .
\end{equation}
Substituting the last result into Eqs.~(\ref{eq:A})-(\ref{eq:F}),
we obtain
\begin{eqnarray}
\label{eq:a-si-A}
 A & =& \frac{1}{\cal D}
  \Big[ s_{11}+s_{11} e^{4ik (\xi_{1} - \xi_{2}) }
  +e^{2ik (\xi_{1} - \xi_{2}) } \Lambda \Big] \\
 \label{eq:a-si-B}
 B & =& \frac{1}{\cal D}
  \Big[ s_{21}+e^{2ik (\xi_{1} - \xi_{2}) }
  \big\{ -s_{21}  (s_{32} s_{23}^{\ast} + s_{33} s_{22}^{\ast}) 
   \nonumber \\
  && + s_{31}  (s_{22} s_{23}^{\ast} + s_{22}^{\ast} s_{23}) \Big] , \\
\label{eq:a-si-C}
 C & =& \frac{1}{\cal D}
  \Big[  (s_{33}^{\ast} s_{21} + s_{23}^{\ast} s_{31}) 
   + s_{11} s_{12}^{\ast} e^{2ik (\xi_{1} - \xi_{2}) } \Big] , \\
\label{eq:a-si-D}
 D & = & \frac{1}{\cal D}
  \Big[ s_{31}+e^{2ik (\xi_{1} - \xi_{2}) }
  \big\{ -s_{31}  (s_{22} s_{33}^{\ast} + s_{23} s_{32}^{\ast}) 
   \nonumber \\
  && + s_{21}  (s_{32} s_{33}^{\ast} + s_{33} s_{32}^{\ast}) \Big] , \\
\label{eq:a-si-E}
 E & =& \frac{e^{2ik (\xi_{1} - \xi_{2}) } }{\cal D} 
  \Big[  (s_{32}^{\ast} s_{21} + s_{22}^{\ast} s_{31}) 
   + s_{11} s_{13}^{\ast} e^{2ik (\xi_{1} - \xi_{2}) } \Big] , \\
\label{eq:a-si-F}
 F & =& \frac{e^{2ik (\xi_{1} - \xi_{2}) } }{\cal D} 
    (s_{31}^{\ast} s_{21} + s_{21}^{\ast} s_{31}) 
    \left(1- e^{2ik (\xi_{1} - \xi_{2}) } \right) , 
\end{eqnarray}
where 
\begin{eqnarray}
 {\cal D} & := & 1- e^{2ik (\xi_{1} - \xi_{2}) }
  \left( s_{22} s_{33}^{\ast} + s_{23} s_{32}^{\ast} 
  + s_{32} s_{23}^{\ast} + s_{33} s_{22}^{\ast} \right)
  \nonumber \\
 &&   + \left| s_{11} \right|^{2} e^{4ik (\xi_{1} - \xi_{2}) } , \\
 \Lambda & := & -s_{11} 
      (s_{22} s_{33}^{\ast} + s_{23} s_{32}^{\ast}
       +s_{32} s_{23}^{\ast} + s_{33} s_{22}^{\ast}) \nonumber \\
 && + s_{12} (s_{33}^{\ast} s_{21} + s_{23}^{\ast} s_{31})
      + s_{13} (s_{32}^{\ast} s_{21} + s_{22}^{\ast} s_{31}) .
\end{eqnarray}
From Eq.~(\ref{eq:a-si-A}), we can find a
condition for perfect transmission ($A=0$).
From $A=0$, we derive
\begin{equation}
 \cos \left( 2k (\xi_{1} - \xi_{2}) \right)
  = - \frac{\Lambda}{2s_{11}} .
\end{equation}
Hence, if the condition
\begin{equation}
 - \frac{\Lambda}{2s_{11}} \in \mathbb{R}
 \quad \mbox{and} \quad 
 \left| \frac{\Lambda}{2s_{11}} \right| \leq 1
\end{equation}
holds, then the perfect transmission occurs repeatedly
as $k$ increases. Furthermore, 
from Eq.~(\ref{eq:a-si-F}),  we find that 
the perfect reflection ($F=0$) occurs if and only if 
the condition
\begin{equation}
 e^{2ik (\xi_{1} - \xi_{2}) } =1
\end{equation}
holds except the trivial cases of $s_{21}=0$ or $s_{31}=0$. 
It should be emphasized that this phenomenon never happen 
when we consider one-dimensional quantum systems with 
double point interactions of degree 2 (see \cite{knt1,knt2}).

\section{Summary}
\label{sec5}

We have newly formulated the system with a single Y-junction
and the ring systems with double Y-junctions. 
For the ring systems, we found the compact formulas for
probability amplitudes based on the path integral approach. 
We have also discussed quantum reflection and 
transmission on the ring systems. Restricting
our attention to the ring systems under the scale-invariant 
junction conditions, we found the conditions for 
perfect transmission and/or perfect reflection. 
It is remarkable that perfect reflection can occur
in the case of the anti-symmetric ring system,
in contrast to the one-dimensional quantum systems
having singular nodes of degree 2.
This phenomena might be related to the possible switching of 
the supercurrent through a ring-shaped Josephson junction, 
which was investigated by \cite{dg}. 
This probable relation would be investigated elsewhere.
General cases beyond the scale-invariant 
junction conditions should also be investigated.
This will be provided in the future works.

%% The Appendices part is started with the command \appendix;
%% appendix sections are then done as normal sections
\appendix

\section{The property of $L_{0}$ in the junction condition}
\label{sec:a-l0}

As shown in \cite{cft}, 
the parameter $L_{0}$ does not provide any additional freedom
for all possible junction conditions characterized by the unitary matrix $U\in U(3)$.
To make the present paper self-contained,
we explicitly show this fact in the case of $U(3)$,
following the discussion of $U(2)$ in \cite{cft}.
When we use $W$ satisfying Eq.~(\ref{eq:diag-w}),
Eq.~(\ref{eq:jc2}) can be written as
\begin{equation}
\label{eq:apd-jc}
 \left( D-I_{3} \right) \tilde{\Psi} 
  + i L_{0} \left( D+I_{3}\right) \tilde{\Psi}' = 0 ,
\end{equation}
where
\begin{equation}
 \tilde{\Psi}= \left( 
 \begin{array}{c}
  \tilde{\Phi}_{1} \\  \tilde{\Phi}_{2} \\ \tilde{\Phi}_{3}
 \end{array}
 \right):=W^{\dagger} \Psi , 
 \quad 
 \tilde{\Psi}' = \left( 
 \begin{array}{c}
  \tilde{\Phi}_{1}' \\  \tilde{\Phi}_{2}' \\ \tilde{\Phi}_{3}'
 \end{array}
 \right):=W^{\dagger} \Psi' .
\end{equation}
Equation (\ref{eq:apd-jc})  is reduced to 
\begin{eqnarray}
\label{eq:apd-jc2}
 \tilde{\Phi}_{j} +  L_{0} \cot \frac{\theta_{j}}{2} \tilde{\Phi}_{j}' =0 
 \quad \left( j=1, 2, 3 \right) .
\end{eqnarray}
We consider an infinitesimal transformation of $L_{0}$ as
\begin{equation}
\label{eq:apd-l0}
 L_{0} \longrightarrow \overline{L}_{0}= L_{0} + \delta L_{0}.
\end{equation}
Under this transformation, if we consider infinitesimal transformations
of $\theta_{j}$  as
\begin{equation}
\label{eq:apd-thj}
 \theta_{j} \longrightarrow \overline{\theta}_{j}
  = \theta_{j}+\frac{\delta L_{0}}{L_{0}} \sin \theta_j.
\end{equation}
then Eq.~(\ref{eq:apd-jc2}) becomes invariant.
Hence, the change of $L_{0}$ can always be absorbed by the
changes of parameters $(\theta_{1}, \theta_{2}, \theta_{3})$. 
In other words, the parameters $(L_{0}, \theta_{1}, \theta_{2}, \theta_{3})$
give the same junction condition as that of 
$(\overline{L}_{0}, \overline{\theta}_{1}, \overline{\theta}_{2}, \overline{\theta}_{3})$.
Thus, the transformation given by (\ref{eq:apd-l0}) and (\ref{eq:apd-thj}) 
can be considered to be a kind of a gauge transformation.
Therefore, we can regard the parameter $L_{0}$ as a gauge freedom.

\section{Gell-Mann Matrices}
\label{sec:a-gm}

The Gell-Mann Matrices are defined as
\begin{equation}
 \lambda_{1} := \left(
  \begin{array}{ccc}
   0 & 1 & 0 \\ 1 & 0 & 0 \\ 0 & 0 & 0 \\
  \end{array} 
 \right) ,
\end{equation}
\begin{equation}
 \lambda_{2} := \left(
  \begin{array}{ccc}
   0 & -i & 0 \\ i & 0 & 0 \\ 0 & 0 & 0 \\
  \end{array} 
 \right) ,
\end{equation}
\begin{equation}
 \lambda_{3} := \left(
  \begin{array}{ccc}
   1 & 0 & 0 \\ 0 & -1 & 0 \\ 0 & 0 & 0 \\
  \end{array} 
 \right) ,
\end{equation}
\begin{equation}
 \lambda_{4} := \left(
  \begin{array}{ccc}
   0 & 0 & 1 \\ 0 & 0 & 0 \\ 1 & 0 & 0 \\
  \end{array} 
 \right) ,
\end{equation}
\begin{equation}
 \lambda_{5} := \left(
  \begin{array}{ccc}
   0 & 0 & -i \\ 0 & 0 & 0 \\ i & 0 & 0 \\
  \end{array} 
 \right) ,
\end{equation}
\begin{equation}
 \lambda_{6} := \left(
  \begin{array}{ccc}
   0 & 0 & 0 \\ 0 & 0 & 1 \\ 0 & 1 & 0 \\
  \end{array} 
 \right) ,
\end{equation}
\begin{equation}
 \lambda_{7} := \left(
  \begin{array}{ccc}
   0 & 0 & 0 \\ 0 & 0 & -i \\ 0 & i & 0 \\
  \end{array} 
 \right) ,
\end{equation}
\begin{equation}
 \lambda_{8} := \frac{1}{\sqrt{3}}\left(
  \begin{array}{ccc}
   1 & 0 & 0 \\ 0 & 1 & 0 \\ 0 & 0 & -2 \\
  \end{array} 
 \right) .
\end{equation}
These eight matrices give bases in SU(3).

\section{The condition of the time-reversal symmetry}
\label{sec:a-tr}

We discuss the condition of the time-reversal symmetry for the 
unitary matrix $U$.
When we consider the time-reversal transformation
\begin{equation}
 t \longrightarrow \overline{t}=-t,
\end{equation}
we find the transformation of the wave function
from Eq.~(\ref{eq:schrodinger}),
\begin{equation}
 \Phi_{i} (t,x_{i} ) \longrightarrow 
 \overline{\Phi}_{i} (\overline{t}, x_{i} ) =\Phi^{\ast}_{i} (t,x_{i} ).
\end{equation}
Then we also have
\begin{equation}
 \Psi  \longrightarrow 
 \overline{\Psi}=\Psi^{\ast}, \quad 
 \Psi'  \longrightarrow 
 \overline{\Psi'}=\Psi'^{\ast}.
\end{equation}
Under the time-reversal transformation, 
the junction condition (\ref{eq:jc2}) is transformed as
\begin{eqnarray}
 \lefteqn{(U-I_{3})\Psi + i L_{0} (U+I_{3}) \Psi' =0 }
  \nonumber \\ 
\label{ep:apd-tr-jc1}
 && \longrightarrow  
 \left( \overline{U}-I_{3} \right)\Psi^{\ast} 
 + i L_{0} \left( \overline{U}+I_{3} \right) \Psi'^{\ast} =0 ,
\end{eqnarray}
where we assumed the transformation
$U \rightarrow  \overline{U}$.
The complex conjugate of Eq.~(\ref{eq:jc2}) 
multiplied by $(-U^{T})$ from the left-hand side, 
where $T$ denotes transposition, becomes 
\begin{equation}
\label{ep:apd-tr-jc2}
 \left( U^{T}-I_{3} \right)\Psi^{\ast} 
 + i L_{0} \left( U^{T}+I_{3} \right) \Psi'^{\ast} =0 .
\end{equation}
Here we have used the property that $U$ is unitary. 
Comparing Eq.~(\ref{ep:apd-tr-jc1}) with (\ref{ep:apd-tr-jc2}),
we derive
\begin{equation}
 U \longrightarrow 
 \overline{U}=U^{T}. 
\end{equation}
Thus, the time-reversal symmetry requires the condition
\begin{equation}
\label{eq:apd-U}
 U  =U^{T}.
\end{equation}
Therefore, if $U$ is a symmetric matrix, then the boundary condition
satisfies the time-reversal symmetry as well as the Schr\"odinger 
equation (\ref{eq:schrodinger}) does.

Next, we discuss the parameter space which satisfies the condition 
(\ref{eq:apd-U}) for the time-reversal symmetry.
From Eqs.~(\ref{eq:u-exp}) and (\ref{eq:apd-U}), we derive
the condition
\begin{equation}
\label{eq:apd-V}
  {\cal V} = {\cal V}^{\ast}, 
\end{equation}
where ${\cal V}$ is given by Eq.~(\ref{eq:V}).
In this case, the S-matrix also becomes 
symmetric, i.e., $S=S^{T}$.
Note that while $i\lambda_{2}$ 
and $i\lambda_{5}$ are real, 
$i\lambda_{3}$ is imaginary. From the requirements
that $e^{i\alpha \lambda_{3}}$, $e^{i\gamma \lambda_{3}}$, and
$e^{ia \lambda_{3}}$ should be real, 
we find the conditions for the time-reversal symmetry,
\begin{eqnarray}
\label{eq:apd-alpha}
 \alpha = 0 \quad  \mbox{or} \quad  \pi, \\
\label{eq:apd-gamma}
 \gamma = 0 \quad  \mbox{or} \quad  \pi, \\
\label{eq:apd-a}
 a = 0 \quad  \mbox{or} \quad  \pi.
\end{eqnarray}
Therefore, when the above conditions 
(\ref{eq:apd-alpha})-(\ref{eq:apd-a}) hold,
the Y-junction satisfies
the time-reversal symmetry, in which 
the six remaining parameters 
$\beta, \delta, b, \theta_{1}, \theta_{2}, \theta_{3}$
are still free.

Finally, we relate our expression of the S-matrix
with the symmetric S-matrix in the previous works \cite{bia,buttiker}.
If we adopt the parameters 
\begin{eqnarray}
 & \alpha =0, \ \beta = \frac{3\pi}{2}, \ 
 \gamma = \pi,\ \delta = \frac{\pi}{4}, \nonumber \\
 & a=0, \ \theta_{1} = 0 ,\ 
 \theta_{2}=\theta_{3}=\pi , 
\end{eqnarray}
then we obtain
\begin{equation}
 S=\left( 
 \begin{array}{ccc}
  -\cos 2b & \frac{1}{\sqrt{2}} \sin 2b &
   \frac{1}{\sqrt{2}} \sin 2b \\
  \frac{1}{\sqrt{2}} \sin 2b & \frac{1}{2} (\cos 2b -1) 
   & \frac{1}{2} (\cos 2b + 1) \\
  \frac{1}{\sqrt{2}} \sin 2b & \frac{1}{2} (\cos 2b + 1)
   & \frac{1}{2} (\cos 2b - 1)
 \end{array}
 \right) .
\end{equation}
It should be noted that this matrix also 
corresponds to a scale-invariant junction condition.
When $0\leq b \leq \frac{\pi}{4}$, 
we introduce new variables
\begin{eqnarray}
 \tilde{\epsilon} & := & \frac{1}{2} \sin^2 2b , \\
 \tilde{a} & := & \frac{1}{2} \left( \sqrt{1-2\tilde{\epsilon}} -1 \right)
  =\frac{1}{2} \left( \cos 2b -1 \right) , \\
 \tilde{b} & := & \frac{1}{2} \left( \sqrt{1-2\tilde{\epsilon}} +1 \right)
  =\frac{1}{2} \left( \cos 2b + 1 \right) . 
\end{eqnarray}
Then we find that the S-matrix is reduced to
\begin{equation}
\label{eq:apd-s-buttiker}
 S=\left( 
 \begin{array}{ccc}
  -( \tilde{a} + \tilde{b} ) & \tilde{\epsilon}^{1/2} &
   \tilde{\epsilon}^{1/2} \\
  \tilde{\epsilon}^{1/2} & \tilde{a}
   & \tilde{b} \\
  \tilde{\epsilon}^{1/2} & \tilde{b}
   & \tilde{a}
 \end{array}
 \right) .
\end{equation}
This is the same expression as the S-matrix in \cite{bia,buttiker}.

\section{The solution of the amplitude
         $A$, $B$, $C$, $D$, $E$ and $F$}
\label{sec:s-a}

By solving Eqs.~(\ref{eq:c-r1}) and (\ref{eq:c-r2}) algebraically, 
we obtain the solution 
for $A, B, C, D, E$ and $F$ as 
\begin{eqnarray}
 A&=& s_{11} + \frac{1}{\Delta}
  \left\{ s_{12} \left( A_{12}(1-A_{33}) 
  +A_{13} A_{32}\right)\right. \nonumber \\
 && \left. + s_{13} \left( A_{13}(1-A_{22}) 
  +A_{12} A_{23}\right) \right\} , \\
 B &=& \frac{1}{\Delta}
  \left\{ s_{31} B_{32} 
   + s_{21} (1-B_{33}) \right\} , \\
 C &=& \frac{1}{\Delta}
  \left\{ A_{12} (1-A_{33}) 
   + A_{13} A_{32} \right\} , \\
 D &=& \frac{1}{\Delta}
  \left\{ s_{21} B_{23}
   + s_{31} (1-B_{22}) \right\} , \\
 E &=& \frac{1}{\Delta}
  \left\{ A_{13} (1-A_{22}) 
   + A_{12} A_{23} \right\} , \\
 F&=& \frac{1}{\Delta}
  \left\{ \tilde{s}_{12} \left( s_{31}B_{32} 
  +s_{21} (1-B_{33})\right)\right. \nonumber \\
 && \left. + \tilde{s}_{13} \left( s_{21} B_{23}  
  +s_{31} (1-B_{22})\right) \right\} , 
\end{eqnarray}
where
\begin{eqnarray}
 A_{ij} & := & \sum_{k=2,3} s_{ki} \tilde{s}_{jk} , \\
 B_{ij} & := & \sum_{k=2,3} \tilde{s}_{ki} s_{jk} , \\
 \Delta & := & (1-A_{22})(1-A_{33})-A_{23} A_{32} \nonumber \\
  &= &  (1-B_{22})(1-B_{33})-B_{23} B_{32} .
\end{eqnarray}

\section{Scale-invariant junction conditions}
\label{sec:apd-sij}

We discuss the Weyl scaling transformation of the wave function
(see also \cite{cft}), which is given by
\begin{equation}
 \Phi_{i} (t, x_{i}) \longrightarrow 
 \overline{\Phi}_{i} (t, x_{i}) 
 = {\cal A} \Phi_{i} (t, \lambda x_{i}) ,
\end{equation}
where $\lambda$ is a positive constant, and 
${\cal A}$ is a normalization factor.
Note that the condition for the normalization of the probability
is provided by
\begin{equation}
 \sum_{i=1}^{3} \int^{\xi}_{-\infty} 
 \left| \Phi_{i} (t, x_{i}) \right|^2 dx_{i} =1 
\end{equation}
in a system with a single Y-junction (Fig.~\ref{fig1}(a)). 
This normalization condition 
becomes, under the Weyl scaling transformation,
\begin{eqnarray}
 \lefteqn{\sum_{i=1}^{3} \int^{\xi /\lambda }_{-\infty} 
 \left| {\cal A} \Phi_{i} (t, \lambda x_{i}) \right|^2 
 dx_{i} } \nonumber \\
 && = \sum_{i=1}^{3} \int^{\xi }_{-\infty} 
 \left| {\cal A} \Phi_{i} (t, y_{i}) \right|^2 
 \frac{1}{\lambda} dy_{i} = 1 .
\end{eqnarray}
Hence, we derive
\begin{equation}
 \Phi_{i} (t, x_{i}) \longrightarrow 
 \overline{\Phi}_{i} (t, x_{i}) 
 = \lambda^{1/2} \Phi_{i} (t, \lambda x_{i}) .
\end{equation}
We also have
\begin{equation}
 \Psi \longrightarrow \lambda^{1/2} \Psi, \quad 
 \Psi' \longrightarrow \lambda^{3/2} \Psi.
\end{equation}
From these results, we find that the junction condition
(\ref{eq:jc2}) is transformed, under the Weyl scaling transformation,
to
\begin{equation}
 \lambda^{1/2} (U-I_{3}) \Psi
 +i L_0 \lambda^{3/2} (U+I_{3}) \Psi' = 0 .
\end{equation}
It follows that the junction condition is not 
invariant in general under the Weyl scaling transformation.
If we impose the Weyl scaling invariance, then 
we obtain 
\begin{equation}
 (U-I_{3}) \Psi = 0, \quad 
 (U+I_{3}) \Psi' = 0 .
\end{equation} 
The last equation means that the eigen values of $U$
take $+1$ or $-1$. In this case, we call the junction condition
scale invariant.

%% If you have bibdatabase file and want bibtex to generate the
%% bibitems, please use
%%
%%  \bibliographystyle{elsarticle-num} 
%%  \bibliography{<your bibdatabase>}

%% else use the following coding to input the bibitems directly in the
%% TeX file.

\end{document}